# Device Modeling Bias in ReRAM-based Neural Network Simulations

Osama Yousuf, Imtiaz Hossen, Matthew W. Daniels, Martin Lueker-Boden, Andrew Dienstfrey, Gina C. Adam

*Abstract*—Data-driven modeling approaches such as jump tables are promising techniques to model populations of resistive random-access memory (ReRAM) or other emerging memory devices for hardware neural network simulations. As these tables rely on data interpolation, this work explores the open questions about their fidelity in relation to the stochastic device behavior they model. We study how various jump table device models impact the attained network performance estimates, a concept we define as modeling bias. Two methods of jump table device modeling – binning and Optuna-optimized binning are explored using synthetic data with known distributions for benchmarking purposes, as well as experimental data obtained from $TiO_x$ ReRAM devices. Results on a multi-layer perceptron trained on MNIST show that device models based on binning can behave unpredictably particularly at low number of points in the device dataset, sometimes over-promising, sometimes under-promising target network accuracy. This paper also proposes device level metrics that indicate similar trends with the modeling bias metric at the network level. The proposed approach opens the possibility for future investigations into statistical device models with better performance, as well as experimentally verified modeling bias in different in-memory computing and neural network architectures.

## I. INTRODUCTION

ARTIFICIAL INTELLIGENCE is exploding and is resource-hungry, with advanced neural networks requiring hundreds of thousands of computing chips. At such large scales, traditional computing cores exhibit excessively high data movement and energy demands owing to the von-Neumann bottleneck, bringing us closer to hitting the "power wall" [1]. Thus, there is a growing need for investigating technologies with non-von Neumann architectures. Such alternatives can serve as building blocks for faster and more efficient neural network accelerators that exploit near- or in-memory computation to minimize data movement. Non-volatile resistive switching devices promise both dense storage and energy-efficient analog processing, making them suitable for in-memory computing for artificial intelligence applications [2], [3].

Hardware-based neural networks based on emerging non-volatile memory devices can exploit underlying physical phenomena to efficiently implement matrix-vector multiplication – a critical operation in neural networks. In particular, ReRAM or memristor devices can be used as highly energy-efficient physical implementations of artificial synaptic weights for neural networks owing to their non-volatility and fast switching characteristics [4], [5]. However, these devices exhibit complex multi-physics behavior leading to performance degradation in prototype networks. As with many other emerging technologies, the impact of these device non-idealities on the network performance needs to be further studied and resolved before they can be used to realize large-scale analog accelerators.

Despite significant progress in the past two decades [4], the study of hardware neural networks based on ReRAM devices is facing several major barriers. First, a purely experimental approach is unfeasible since commercial ReRAM tape-outs have long timelines and significant design and fabrication costs [6]–[8]. Secondly, before hardware prototyping can be practically motivated, results from hardware-aware simulations are needed to reliably predict the promising performance of these systems and optimize their end-to-end behavior across all levels – from underlying devices to neural networks. Therefore, suitable modeling of ReRAM devices is a key consideration for hardware neural networks investigations and in-memory computing systems based on ReRAM devices in general.

A broad range of models has been proposed. Atomistic models focus entirely on simulations from first principles in an effort to capture the multi-physics dynamics of filament formation and their conductance characteristics. These models may be used to provide insights needed for material and device design optimization [9]–[12]. Compact models needed for circuit design use physically inspired parametrizations which are tuned to match experimental current *vs.* voltage characteristics [13], [14]. While both classes of models are appropriate for their respective purposes, they rely on computationally expensive systems of equations and are therefore not well-suited for scaling to neural network simulations. Additionally, these models require investigating

This work was supported in part by NIST under grant 70NANB22H018, by Western Digital under grant ECNS21932N and the GW Cross-Disciplinary Research Fund. Corresponding author: Gina C. Adam (e-mail ginaadam@gwu.edu).

Osama Yousuf, Imtiaz Hossen, and Gina C. Adam are with the George Washington University, Electrical and Computer Engineering Department, Washington, DC, 20052. Matthew W. Daniels is with the National Institute of Standards and Technology, Gaithersburg, MD, 20899, USA and Andrew Dienstfrey is with the National Institute of Standards and Technology, Boulder, CO, 80305 USA. Martin Lueker-Boden is with the Western Digital Technologies, San Jose, CA, 95119, USA.

underlying physical phenomena such as the shape of the filament [15]–[17] for new devices, which is a time-consuming process that can delay algorithmic investigation.

Due to these limitations, for the following study we propose to investigate jump table models to characterize ReRAM device irregularities. Such models consist of lookup tables that specify the probability of moving from one conductance state to another as a function of present conductance state and applied pulse (and potentially other measurement parameters). In contrast to the physics-based models discussed above, these models are derived solely from experimental data without reference to underlying physical mechanisms. While these physical mechanisms are important for device design, they are not relevant in predicting neural network performance. It is known that variability in the switching of ReRAM devices leads to weight dispersion, which can potentially negatively impact the accuracy and performance of the overall system [18], [19]. Therefore, accurate device modeling is needed for providing a realistic estimate of the training characteristics of analog neuromorphic systems implemented with real ReRAM devices.

In this paper, we introduce and study *modeling bias* as a useful concept to evaluate the fitness of a device model in the context of neural network simulations. We investigate the impact of two methods for creating jump table-based device models: the traditional binning and interpolation method proposed in [20], and an iterative hyperparameter optimization method based on the Optuna framework [21]. For benchmarking purposes, we synthesize ReRAM jump tables using an assumed closed-form model for device switching and investigate the convergence of our interpolation methods, as detailed in Section II A. These distributions are inspired by the "Real Device" model from the NeuroSimV3.0 simulator [22]. Trends in network metrics are compared with trends in the goodness-of-fit for the device to systematically analyze the performance of the interpolation methods for device modeling in the context of neural network simulations. We also present the training performance and methods to estimate the modeling bias when *experimental jump tables* are used. These are jump tables based on experimental data obtained from $TiO_x$ ReRAM devices, for which the underlying distributions for signal mean and switching noise are not known.

The remainder of this paper is organized as follows. Section II covers the methods used in our analysis including details on jump table modeling, device datasets, neural network structure and training routines, and metrics for inferring data interpolation quality. Section III presents comparative neural network results using (a) synthetic datasets, where we investigate the impacts of the number of input data points, device cycle-to-cycle variation, and device non-linearity on modeling bias, and (b) experimental datasets, where we present an iterative approach for jump table modeling based on hyperparameter optimization and show how the modeling bias performance could potentially be estimated without experimental apparatus. Section IV is a discussion highlighting the limitations and opportunities in emerging device modeling for large-scale neural network simulations. We wrap up with conclusions in Section V.

## II. METHODS

### A. Jump table modeling

A jump table is a set of cumulative distribution functions (CDFs) that define the stochastic change in device conductance per voltage pulse as a function of the current conductance state (i.e., the distribution of ΔG per pulse as a function of G) [18], [20], [23]. We define a jump table device model $x$ as a series of distributions as follows:

$$x = \{X_{G_1}, X_{G_2}, \ldots, X_{G_m}\}, \qquad (1)$$

where each $X_{G_i}$ is a normal distribution with mean $\mu_i$ and standard deviation $\sigma_i$, and m is the total number of distributions. The *mean profile* $\mu(G)$ refers to the ordered list of $\mu_i$'s, and likewise, the *standard deviation profile* $\sigma(G)$ for $\sigma_i$'s.

Jump tables provide a single model to encompass both the stochastic nature of device programming and the non-uniform conductance response ΔG as a function of conductance G. The cycle-to-cycle variability of a device is represented by the standard deviation around the ΔG mean. An alternative representation in the resistance space (ΔR vs. R) [24] is also possible. Here, we opt to work in (G, ΔG)-space since conductances map naturally to matrix weights in a neural network implemented with real ReRAM devices.

During training, a desired weight update is determined by the gradient of the loss, and programming pulses are applied to each device to adjust its conductance according to this update [18]. Due to ReRAM physics, conductance increases and decreases are not always symmetrical. Thus, two jump tables are needed – one for potentiation (increase in G), and another for depression (decrease in G), referred to as the SET and RESET tables respectively.

Jump tables are constructed from sampled data. For example, a dataset could consist of ReRAM measurements of n experimentally sampled points $(G_1, \Delta G_1), \ldots, (G_n, \Delta G_n)$, where for a given i, $G_i$ is the initial device conductance and $\Delta G_i$ is the conductance change resulting from a single pulse at a fixed voltage. Alternatively, one may construct a synthetic dataset by sampling an arbitrary number of $(G_i, \Delta G_i)$ data points from a given mean and standard deviation profile. Given this discrete data, various interpolation methods can be used to create a continuous model of the underlying mean and standard deviation profiles.

The binning jump table device modelling algorithm is outlined in Algorithm 1. The inputs are a dataset D comprising of $(G_i, \Delta G_i)$ pairs, and the number of bins $G_{bins}, \Delta G_{bins}$. The outputs are the binning model's mean and standard deviation profiles $\mu(G)$ and $\sigma(G)$. In brief, the $(G_i, \Delta G_i)$ data is grouped into bins of specified count. From each G bin in this binned data, the mean and standard deviation information for the $\Delta G_i$ are computed. The mean and standard deviation values are then linearly interpolated across the G bins to obtain the jump table model.

It has been observed that binning and interpolation can introduce unwanted errors and artifacts in modeling such as edge effects, empty bins, excessive smoothing, and rugged predictions of the underlying mean and standard deviation [25]. To avoid such artifacts, parameters $G_{bins}$ and $\Delta G_{bins}$ for Algorithm 1 are optimized for each input dataset individually, in line with our previous work [26]. This is done to ensure that comparisons between different binning models remain fair relative to one another.

---

**Algorithm 1.** Binning Jump Table Device Modelling

| | |
|---|---|
| 1 | **procedure** BinningModel (**in** D, $G_{bins}$, $\Delta G_{bins}$; **out** μ, σ) |
| 2 | hist, edges = **Hist2D**(D, $G_{bins}$, $\Delta G_{bins}$)   // 2D binning |
| 3 | $G_{axis}$, $\Delta G_{axis}$ = **GetMean**(edges) // average G, ΔG per bin |
| 4 | C = [] |
| 5 | **for** i in **range**(hist) **do** |
| 6 | C[i] = **ToCDF**(hist[i])   // calculate CDF(ΔG) per G bin |
| 7 | end **for** |
| 8 | μ, σ = **fit**(C, $G_{axis}$, $\Delta G_{axis}$)    // Gaussian fit, interpolated |
| 9 | end **procedure** |

---

Details of the Optuna model are presented separately in Section II E. To summarize, both the binning and Optuna modelling schemes attempt to predict the underlying mean and standard deviation profile of the input conductance data based on data interpolation.

*B. Device Datasets*

Cycle-to-cycle switching variability present in ReRAM devices is a non-ideality that can significantly hamper the performance of neural networks based on ReRAM devices [27]. Significant efforts in the community focus on engineering devices with lower variability [28] and providing algorithmic approaches to manage this device variability at the neural network level [20], [29]. For this work, we are focused only on exploring the impact of cycle-to-cycle variability on the modeling bias at the network level. Different formalisms exist in the literature to quantify device non-idealities. For example, [22] expresses cycle-to-cycle switching variation as an overall percentage of the underlying conductance range, [30] expresses this variability in terms of the ratio between the standard deviation and device resistance, and [31] proposes the use of a stochastic translator which uses a bitwise-AND operation in place of multiplication to capture this switching noise. We model the cycle-to-cycle variability by a random variable with distribution given by the CDF within a jump table.

The underlying distribution and noise profile are unknown in experimental data. Therefore, for the purposes of testing the proposed methodology, we use synthetic target distributions. Synthetic Gaussian datasets are generated with pre-determined underlying mean and standard deviation profiles μ(G) and σ(G) over a given conductance range $G \in [G_{min}, G_{max}]$ with fixed minimum and maximum bounds. A synthetic jump table can be created either from these target mean and standard deviation profiles, called a *target jump table*, or from interpolated predictions of these profiles obtained through the binning model, referred to as a *binning jump table*.

In this work, we focus on a family of synthetic distributions for which the random variable representing ΔG per pulse has a mean profile μ(G) that is linear in G, and the standard deviation profile σ(G) is a constant. This was chosen to match the experimentally derived "*Real Device*" model from the MLP+NeuroSim V3.0 framework [22], which we refer to as the *analytical device model* hereafter.

The jump table for this analytical model was synthesized using the theoretical equations from NeuroSim; we first set the cycle-to-cycle variation (and all other device variations) to zero, keeping only the non-linearity values for potentiation and depression as 2.4 and −4.88 respectively (which are the default values for Ag:Si-based ReRAM model in NeuroSim). The maximum number of conductance levels was set to 500. We use this model to extract ($G_i$, $\Delta G_i$) datapoints for potentiation as well as depression. As briefly already highlighted, the mean of this device model followed a linear profile, with a constant standard deviation profile. Through data fitting, it was found that the cycle-to-cycle variation parameter of the analytical model, C2C, related to the constant standard deviation profile, σ, of the jump table model as follows:

$$\sigma(G) = C2C \cdot (G_{max} - G_{min}), \quad (2)$$

where $G_{min}$ and $G_{max}$ are physical minimum and maximum bounds on the device conductance values, fixed at 3 nS and 38 nS respectively. In essence, the standard deviation relates to the cycle-to-cycle switching variation as a percentage of the conductance switching range.

With the extracted mean profile μ(G) and the standard deviation σ(G) using (2), we can sample ΔG vs. G datasets for any given analytical device model containing an arbitrary number of points at a specific cycle-to-cycle variation.

---

**Algorithm 2.** Synthetic Data Generation

| | |
|---|---|
| 1 | **procedure** GenerateSyntheticData (**in** n, G, μ, σ; **out** D) |
| 2 | f = **LinearInterpolate**(**x** = G, **y** = μ) |
| 3 | g = **LinearInterpolate** (**x** = G, **y** = σ) |
| 4 | **for** i in range [1, n] **do** |
| 5 | $G_i$ = **RandomUniformFloat**(min(G), max(G)) |
| 6 | $\mu_i$ = **f**($G_i$), $\sigma_i$ = **g**($G_i$) |
| 7 | $\Delta G_i$ = **RandomNormalFloat**($\mu_i$, $\sigma_i^2$) satisfying (3) |
| 8 | D[i] = ($G_i$, $\Delta G_i$) |
| 9 | end **for** |
| 10 | end **procedure** |

---

The synthetic device data generation algorithm used is outlined in Algorithm 2. The inputs are the number of points to synthesize n, conductance dynamic range G, and the mean and standard deviation profiles μ(G) and σ(G). The output is a synthetic dataset D comprising ($G_i$, $\Delta G_i$) pairs. Due to physical device constraints, for all pairs ($G_i$, $\Delta G_i$) recorded experimentally, we also require that

$$G_{min} \leq G_i + \Delta G_i \leq G_{max}, \quad (3)$$

that is, the jump table may not push a device beyond its minimal or maximal conductances. For any conductance update that exceeded these bounds, the conductance is clipped to the corresponding extreme value in order to eliminate non-physical device behavior. Additional constraints may be imposed to further reflect physical realism.

Fig. 1a depicts the one-to-one matching between the conductance trajectories for both SET and RESET phases of the analytical model and the corresponding target jump table model. The Ag:Si analytical ReRAM model from NeuroSim has a cycle-to-cycle variation of 3.5 %, whereas our equivalent target jump table model has a standard deviation of 1.2 nS. Fig. 1b shows the overlapping raw conductance data of the two models, indicating that our synthetic model is able to generate

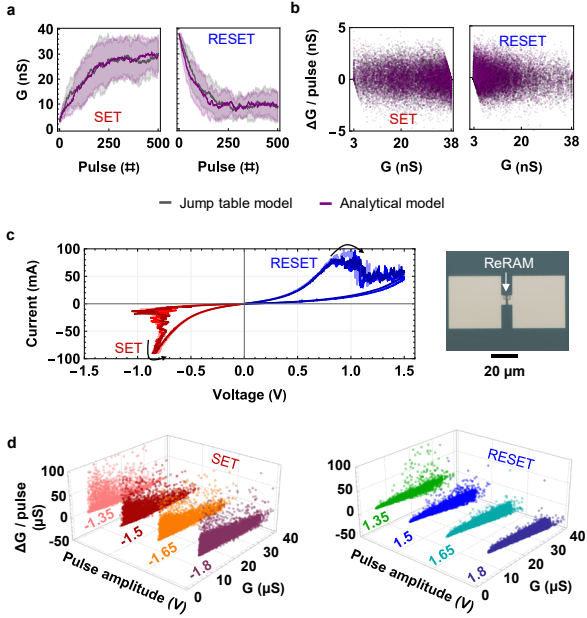

**Fig. 1.** Overview of device data used in our modelling. *Synthetic:* (a) Device conductance trajectories and (b) associated ΔG vs. G data of the analytical and target jump table device models at 3.5 % cycle-to-cycle variability and 1.2 nS standard deviation respectively. *Experimental:* (c) ReRAM devices used for experimental data gathering showing I-V characteristics, and (d) ΔG vs. G data at different pulse amplitudes.

($G_i$, $\Delta G_i$) data points in a dynamic range which matches the analytical model.

In this work, we also use experimental ΔG vs. G data from a 10 nm device with 2.5 nm $Al_2O_3$ / 15 nm $TiO_x$ / 5 nm Ti / 30 nm Pt. The data were obtained utilizing a semiconductor parameter analyzer and pulse programming (Fig. 1c). Before programming, the devices were formed with monotonically increasing voltage pulses to reach a high conductance state. The data were collected after forming and then cycling between SET and RESET 30 times. A device is first programmed to a random conductance value within a given range. Then a write pulse voltage with 500 ns high time, 100 ns rise and fall time, and voltage amplitude randomly selected from among {±1.35 V, ±1.5 V, ±1.65 V, ±1.8 V} is applied. The conductance state is read via a subsequent pulse of 100 mV applied 100 μs after the write pulse; during the intermediary time, the device is held at 0 V. This process is repeated until either all pulses in the list have been applied or the device conductance exceeds some defined limit, then the algorithm restarts from a random conductance value. The end result is a separate (G, ΔG) dataset for every voltage amplitude (Fig. 1d). The exploration of pulse amplitudes is purposely random: typical approaches that use monotonically increasing or decreasing write pulse voltage steps may result in sparse data sets for high write voltages, since the device typically switches before the higher voltages are applied. By comparison, our resulting data sets for each write pulse voltage have roughly the same number of points (≈10,000). This is critical for the jump table modeling of this device data.

*C. Neural Network Details*

A 2-layer perceptron network with 324 neurons in the input layer, 50 neurons in the hidden layer, and 10 neurons in the output layer, with no biases, was used. The network was trained for image classification on a reduced version of the MNIST dataset, which consists of 20x20 pixel images (center-cropped from the original 28x28 pixel images) of handwritten digits labelled 0 through 9 [32]. Inputs to the network were first normalized by subtracting a constant (0.1307) and dividing the difference by a fixed scale factor (0.3801). This shift and scaling were determined as the mean and standard deviation of all pixels in the 60,000 images of the original, uncropped MNIST training set. A sigmoidal activation function was used for each hidden neuron, and a softmax activation was used for the output layer. Mean squared error was used as the loss function for training. The weights were initialized uniformly at random from $-\sqrt{k}$ to $\sqrt{k}$, where $k$ is the reciprocal of the number of input features for a given layer (thus $k = \frac{1}{400}$ and $k = \frac{1}{10}$ for the input and output layers respectively.) Mini-batch gradient descent was used based on previous investigations [29], [33], [34] of its performance benefits when training a network with non-ideal synaptic weights, and the batch size was fixed at 4096. Furthermore, all network-level quantities such as layer weights, gradients, activations and errors were quantized to 6-bit fixed-point precision using the QPyTorch library [35]. The choice of network architecture, loss function, optimizer, and the quantization scheme are all motivated by the specifications of our future hardware prototype [36].

Fig. 2 presents background on how a feed forward neural network can be implemented using ReRAM devices in hardware, and on their modelling via jump tables. A single ReRAM device behaves like an artificial synapse, and ReRAM device crossbars can physically implement densely connected synaptic weight matrices, with weights $w_{ij}$ of a network layer represented as device conductances $G_{ij}$ in the memristive crossbar array. For inference, pixels of an input image are converted to voltages on the input wires of the ReRAM crossbar, as shown in Fig. 2a. As a result of resistive circuit physics (Ohm's and Kirchhoff's Laws), these voltages induce currents on the layer-output wires whose values are equal to the matrix-vector product of synaptic weights times input voltages. These currents are then converted into numerical values using an analog-to-digital conversion. The nonlinear activation function is computed using digital logic. This result is fed as a voltage applied to the inputs of the following network layer, and so on until the final output layer is reached.

For backpropagation, traditional minibatch gradient descent can be used to update the conductances of the crossbar array. For a conductance $G_{ij}$ at minibatch iteration k, firstly the total number of pulses $p_{ij}$ to be applied is determined by:

$$p_{ij} = p_{max} \cdot \left(\nabla_w^{(k)} L\right)_{ij}, \qquad (4)$$

where $\nabla_w^{(k)} L$ is the gradient of the loss function averaged over all minibatch inputs at iteration k, and $p_{max}$ is the device dynamic range, defined as the total number of pulses required to either SET the device from $G_{min}$ to $G_{max}$, or alternatively, RESET the device from $G_{max}$ to $G_{min}$. In hardware, the $p_{ij}$ pulses would physically be applied to the conductance $G_{ij}$ directly. However, in our jump table training regime, the following algorithm is employed to update $G_{ij}$ due to the train of applied pulses $p_{ij}$: A random number $u_k$ is generated between 0 and 1. The corresponding conductance change for a pulse is the inverse of the CDF for a device at conductance $G_{ij}$ applied to u, $\Delta G = CDF^{-1}(u)$. The device conductance is updated to $G_{ij} + \Delta G$ and this cycle is repeated for all pulses. These jump table conductance updates intrinsically quantize and model the conductance update behavior of real devices.

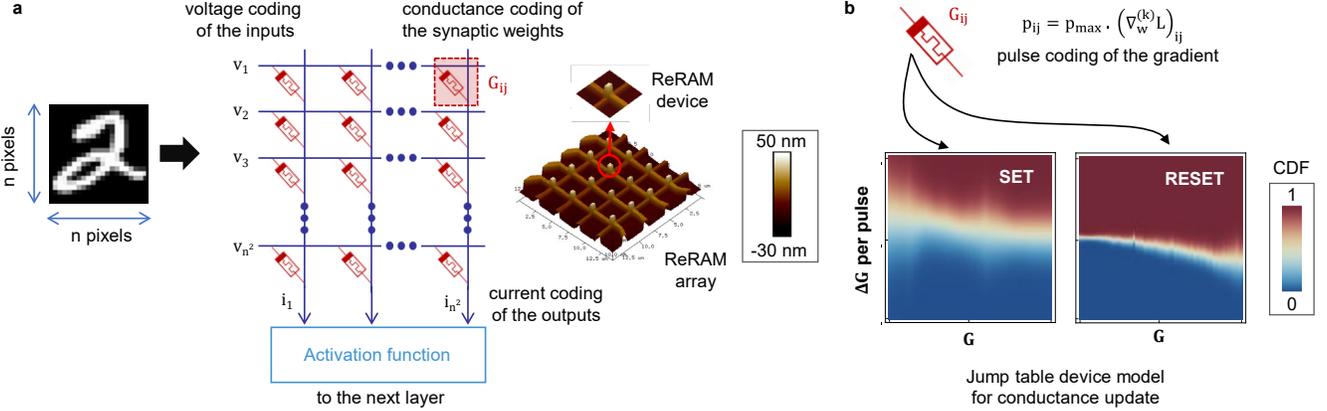

**Fig. 2.** Inference and training in a ReRAM-based multi-layer perceptron network. (a) *Inference:* ReRAM conductances implement matrix-vector multiplication in hardware as a consequence of electrical circuit physics (Ohm's and Kirchoff's Laws). Atomic Force Microscopy picture of a representative crossbar matrix of individual ReRAM devices is also shown. (b) *Training:* For all ReRAM devices in the network, the derivative of the loss function with respect to the synapse is converted to the number of voltage pulses to be applied to the corresponding device. The change in conductance resulting from these pulses is modelled by two jump tables, one for SET and one for RESET, samples of which are also shown.

We implemented jump table-based neural network training in PyTorch [37]. In our simulations, the synaptic weights for each network layer are represented in the conductance space as a crossbar array via the CDF given by a pair of SET and RESET jump tables (Fig. 2b). While the scope of our work is modeling neural networks based on ReRAM devices, our framework can easily be extended to model other emerging analog memory technologies such as phase change memories, magnetic tunnel junctions, or photonic modulators [18], [38]. Additionally, this work is entirely focused on training and investigating modeling bias using device models during training. Once programmed, the ReRAM device weight behaves like a fixed and relatively stable circuit element, so any small read noise and relaxation effects were considered negligible for the inference phase.

For all target models, a learning rate optimization was implemented using a grid-search. In brief, training was carried out using learning rates sampled on a logarithmic scale covering a wide range. The optimal learning rate was defined as that which gave the highest test accuracy averaged over the first ten training epochs. Subsequent network training simulations were computed over 100 epochs with the learning rate set to this value in order to isolate the effects of jump table models on network performance.

*D. Metrics for Evaluating Modeling Quality*

In order to quantify the quality of the device model in the context of the neural network training, the following metrics are employed at the network and device levels.

*D.1. Modeling Bias*

Given that ReRAM conductance updates have a stochastic character, any single training simulation and its attendant accuracy should be viewed as an instance of a random variable. Different models for the ReRAM devices have different distributions in accuracy, and it is sensible to compare their means. We define modeling bias (MB) as the amount that the network's average modelled test accuracy (Mean Acc.$_{model}$) – the accuracy when the synaptic weights are represented by ReRAM devices using an interpolated jump table model – differs from the average target test accuracy (Mean Acc.$_{target}$). These averages are computed for each epoch as shown in (5). Note that the target test accuracy can have a different meaning based on the type of data we used for modeling. For synthetic data, it is the accuracy when the synaptic weights are modelled by ReRAM devices using the synthesized target jump table model. For experimental data, it is the experimental test accuracy that would be obtained in a physical implementation based on the real behavior of the device population used for creating the jump table models.

Since modeling bias is a function of the training epoch, a good device model is one that has lower modeling bias throughout training. Formally, we define

$$MB(i) = \text{Mean Acc.}_{model}(i) - \text{Mean Acc.}_{target}(i) \quad (5)$$

as a function of epoch number i. The bias can either be positive, when the modelled accuracy is too optimistic and overpredicts the target accuracy, or negative, when the modelled accuracy underpredicts it. For the results shown below, mean network accuracies required by (5) were determined by averaging over 20 instances of network training for a given device model.

*D.2. Switching Sign Discrepancy and Overlapping Error*

To estimate the quality of a device model, we propose the use of two device-level error metrics: switching sign discrepancy (SSD) and overlapping error (OVLE). At the network level, the switching direction of ReRAM devices plays a pivotal role in effective training. As an example, suppose the target mean values for a given jump table model were all positive across the entire range of the G parameter with minimal variance, as is the case with some of our SET tables. If an interpolated model regularly predicts negative $\Delta G$ updates for many of the corresponding G indices, the result would be a weight that is changed in the opposite direction of what is prescribed by the corresponding gradient. These updates with inverted signs would almost certainly lead to a degradation in training performance. The same argument applies to the RESET tables. To measure the presence of this failure mechanism, we define the switching sign discrepancy (SSD) as:

$$\text{SSD}(x, y) = \frac{1}{n}\sum_{i=1}^{n} |P(X_{G_i} < 0) - P(Y_{G_i} < 0)|, \quad (6)$$

where $X_{G_i}$ and $Y_{G_i}$ are distributions of device models x and y respectively, n is the total number of distributions within these models (assumed to be the same for both models), and $P(X_{G_i} < 0)$ is the probability that the i-th distribution of model x produces a negative variate. For our jump table models with Gaussian distributions, this is computed as:

$$P(X_{G_i} < 0) = \int_{-\infty}^{0} \text{PDF}_{X_{G_i}}\, dG = \frac{1}{2}\text{erfc}\left(\frac{\mu_i}{\sigma_i\sqrt{2}}\right), \quad (7)$$

where $\text{PDF}_{X_{G_i}}$ is the probability density function of $X_{G_i}$, erfc(.) is the complementary error function, and $\mu_i$, $\sigma_i$ are the mean and standard deviation of $X_{G_i}$. The SSD quantifies the average signed agreement between two device models in terms of their switching behavior. The SSD of a device model with itself is 0 since the agreement is maximum and is 1 when there is no signed switching agreement between two models.

Along similar lines, the overall overlap between device model distributions is also of importance. For this, we make use of the complement of the overlapping coefficient from [39] and define the overlapping error (OVLE) between two device models x and y as follows:

$$\text{OVLE}(x, y) = \frac{1}{2n}\sum_{i=1}^{n} \int_{-\infty}^{+\infty} |(\text{PDF}_{X_{G_i}} - \text{PDF}_{Y_{G_i}})|\, dG. \quad (8)$$

The OVLE of a device model with itself is 0 since the overlap is maximum and is 1 in the case of no overlap between two models.

These device metrics can potentially present insights about the modeling bias performance of a given device model, prior to carrying out expensive neural network experiments, and can thus save significant time. Additionally, these metrics can also serve as a tool for device designers to quickly quantify disagreement (or agreement) between different devices at the level of device switching. As an example, different jump table models could be extracted from experimental $(G_i, \Delta G_i)$ datasets obtained from different devices, and these error metrics could be computed and compared across device populations.

*D.3. Kolmogorov-Smirnov test*

For experimental data where the underlying statistics are unknown, the SSD and OVLE cannot be directly calculated. The two-sample Kolmogorov-Smirnov test (or a similar non-parametric two sample test) can be used instead. The Kolmogorov-Smirnov test can provide a probabilistic estimate of two datasets being drawn from the same underlying distribution. The test value reports the largest absolute difference D between two distribution functions I(x) and J(x) as given by (9).

$$D = \sup_{x} |I(x) - J(x)|, \quad (9)$$

where sup denotes the supremum. A lower test value corresponds to a better fit, meaning that the probability of the two datasets having been drawn from the same distribution is higher. Since our experimental device data is two-dimensional $(G_i, \Delta G_i)$, we use a two-dimensional variant of the Kolmogorov-Smirnov test, as specified in [40]. Our methodology is as follows: First, the experimental data is transformed to reflect a compliance current to limit switching between $G_{min}$ and $G_{max}$ using (3). Second, the data is randomly split into non-overlapping modeling and testing subsets. The modelling subset is used to generate binning and Optuna jump table models. We then use these models to synthesize a $(G_j, \Delta G_j)$ dataset. Finally, we apply the Kolmogorov-Smirnov test on the initial testing subset and each of the two synthesized datasets. As a baseline for comparison, we additionally report the Kolmogorov-Smirnov test values between the two experimental subsets in our results.

*E. Optuna modelling*

We propose a method to explore the network modeling bias in the absence of experimental apparatus and/or experimental network accuracies. Ideally, the experimental data includes a complementary set of experimental device $(G_i, \Delta G_i)$ datapoints and the corresponding network train / test accuracies obtained from a prototype network physically implemented with these devices. In such case, different device modeling methods, their device- and network-level metrics, and respective network simulation results could be tested against the experimental reality directly to determine if the device model needs to be further optimized. However, prototyping neural networks in emerging hardware is challenging, expensive and time consuming. Therefore, for a given emerging device technology, only experimental device $(G_i, \Delta G_i)$ data might be available to benchmark both the device model and the network simulation results. Nevertheless, the question remains if the device model utilized is sufficient to obtain a realistic network accuracy or if there is space for optimization in the device model.

To address this challenge, this paper presents a device model optimization methodology based on Optuna [21], an automated hyperparameter optimization framework. The algorithm employs a derivative-free optimization strategy to push the limits of device modeling towards experimental realism. As an optimization metric, the Kolmogorov-Smirnov test is used. Algorithm 3 details our developed Optuna algorithm. The input is an experimental dataset D comprising of $(G_i, \Delta G_i)$ pairs, conductance range G, polynomial degrees m and n to use for modelling mean and standard deviation profiles respectively, objective function f, and a termination value e to be optimized. In our case, the objective function is the two-dimensional two sample Kolmogorov-Smirnov error between the testing subset of the experimental dataset, and synthesized Optuna dataset. However, this can easily be replaced with an alternate metric as needed. e is chosen as the binning test error, which is the Kolmogorov-Smirnov error between the testing subset of the experimental dataset and a synthesized binning dataset. The outputs are Optuna mean and standard deviation profiles $\mu'(G)$ and $\sigma'(G)$, the test error of which is better than the corresponding binning model by iterative optimization.

| | **Algorithm 3.** Optuna Jump Table Device Modelling |
|---|---|
| 1 | **procedure** OptunaModel (**in** D, G, m, n, f, e; **out** $\mu'$, $\sigma'$) |
| 2 | $e' = \infty$ (Initialize Optuna error as inf.) |
| 3 | $\mu_{params}$ ← Initialize m + 1 coefficients for μ(G) profile |
| 4 | $\sigma_{params}$ ← Initialize n + 1 coefficients for σ(G) profile |
| 5 | **while** $e' \geq e$ **do** |
| 6 | $\mu'$ = **EvaluatePolynomial**(G, $\mu_{params}$) |
| 7 | $\sigma'$ = **EvaluatePolynomial**(G, $\sigma_{params}$) |
| 8 | **GenerateSyntheticData**(**in** length(D),G,$\mu'$,$\sigma'$; **out** D') |
| 9 | $e'$ = **f**(D', D) |
| 10 | $\mu_{params}$ ← Update m + 1 coefficients for μ(G) profile |
| 11 | $\sigma_{params}$ ← Update n + 1 coefficients for σ(G) profile |
| 12 | end **while** |
| 13 | end **procedure** |

In essence, Optuna iteratively learns the mean and standard deviation profiles of an input experimental dataset consisting of $(G_i, \Delta G_i)$ points recorded at a certain voltage by fitting two separate polynomials (of degrees m and n) on the input dataset – one for the mean profile over G, and another for the standard deviation profile over G. The m + 1 coefficients for the mean profile, and the n + 1 coefficients for the standard deviation profile serve as hyperparameters for optimization. Since the algorithm is iterative, there is no guaranteed upper-bound on its time complexity. In comparison with binning, it is fairly more time-consuming in terms of runtime because of having to sample from a mn-dimensional space a large number of times.

For our datasets, hyperparameters were initialized based on direct polynomial fitting over binning mean and standard deviation profiles. Successive sampling was done by the tree-structured parzen estimator – a Bayesian optimization strategy. For all RESET voltage pulse amplitudes from our experimental datasets, the termination criterion was achieved using parabolic modelling of the mean and standard deviation profiles, and for all SET voltages, it was achieved using linear modelling of the mean and standard deviation profiles.

III. RESULTS

A. Impact of number of data points

Identifying the optimal number of (G, ΔG) data points for jump table modeling is of significant interest to provide a highly representative simulation result at the network level, while keeping the device measurement time to a minimum.

To study this, we generated series of jump tables following the linear mean, constant standard deviation profiles from Fig. 2 (equivalent to 3.5% cycle-to-cycle variability of the analytical device model) with an increasing number of points in the raw dataset, ranging from 40 to a total of 10,000 data points. Neural network simulations using these jump table models were run at their respective optimized hyperparameters and the modeling bias as well as SSD and OVLE metrics were determined.

The results in Fig. 3 show a potential correlation between the quality of the device model and the quality of the network simulation based on this device model. As the number of data points increases, conductance trajectories of the binning models align more with the target model (Fig. 3a, b). Additionally, device-level metrics SSD and OVLE of the binning models both decrease, and so does the modeling bias of the corresponding binning jump table models (Fig. 3c, d). This occurs because the binning algorithm is able to more effectively capture the underlying target mean and standard deviation profiles given rich data, leading to overall good network-level estimation of the target models (Fig. 3e).

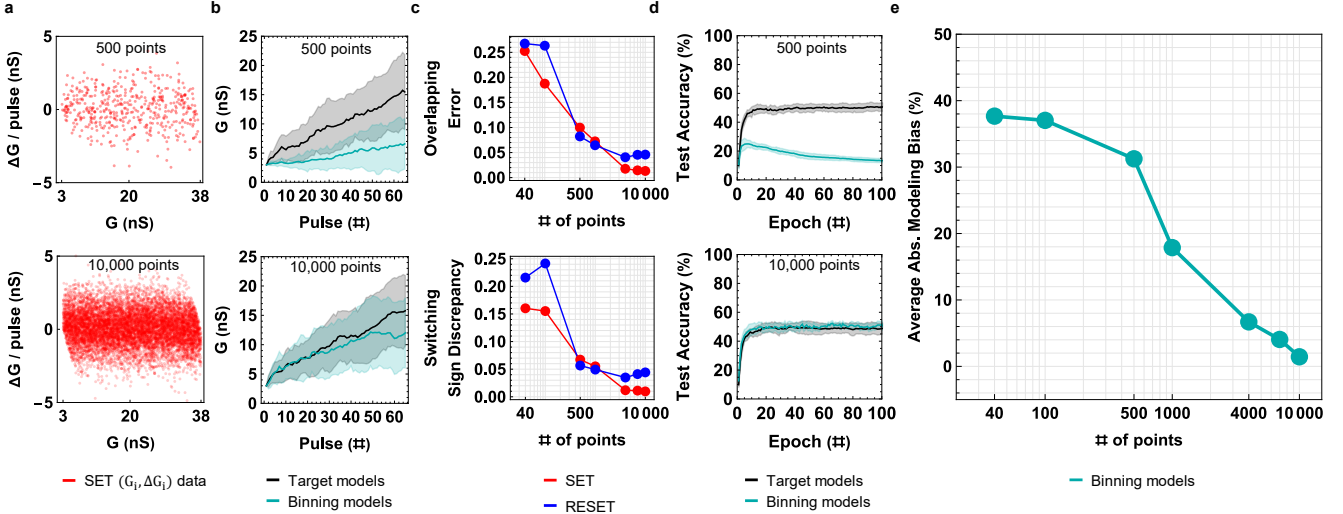

**Fig. 3.** Impact of the number of points for device modelling and respective network simulations. (a) Synthesized (G, ΔG) datasets for SET modelling with 500 and 10,000 points, and (b) corresponding conductance trajectories of the target device model and interpolated binning model for up to 64 applied pulses. (c) Overlapping Error and Switching Sign Discrepancy of different binning jump table models. (d) Network training curves of the 500 point and 10,000-point models, and (e) average absolute modeling bias of the different binning jump table models, indicating that the modeling bias performance improves as the number of points increase. The batch size used is 4096 and optimized network learning rate is 0.1.

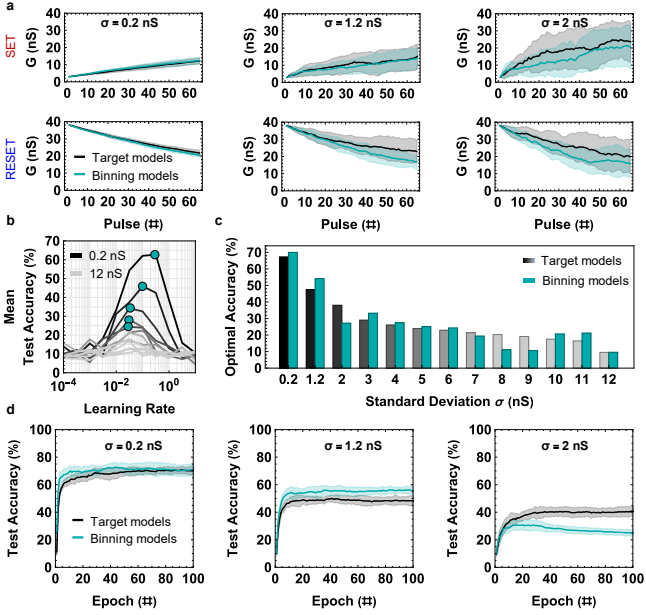

**Fig. 4.** Impact of the cycle-to-cycle variability on device modelling and respective network simulations. (a) Conductance trajectories of target and binning jump table models at different levels of standard deviation for up to 64 applied pulses. (b) Training windows of target models. Cyan markers indicate optimal learning rates. (c) Comparative optimal accuracies of target and binning models. Traces in (b) correspond to target models of the same shade in (c). (d) Convergence curves corresponding to the device models in (a).

There is minimal benefit in terms of modeling bias upon increasing the points from 4,000 to 10,000. The remainder of the jump tables generated in this work are thus based on 4,000 samples of $(G_i, \Delta G_i)$ points, where these samples are drawn from target Gaussian distributions with known mean and standard deviation profiles in the synthetic case and are experimental measurements when working with real devices.

*B. Impact of Cycle-to-Cycle Variability*

To investigate how the device cycle-to-cycle variability impacts the device modeling and network simulation respectively, we synthesized variants based on the analytical device model at increasing values of the underlying standard deviation, sampled in a range from 0.2 nS to 12 nS, corresponding roughly to cycle-to-cycle variations from 0.5 % to 34 % of the underlying conductance range. The mean profile was not altered, and the number of $(G, \Delta G)$ data points was fixed at 4,000.

Fig. 4 presents an analysis of the overall performance of these models. As the cycle-to-cycle variability of target devices increases, the target conductance trajectories become noisier. As a consequence, the corresponding binning models find it harder to predict the underlying mean and standard deviation profiles. This manifests in binning conductance trajectories deviating from target trajectories (Fig. 4a).

At the network level, the increase in cycle-to-cycle variability translates to an overall degradation of the training capability of target devices, evident from narrower learning rate training windows in Fig. 4b and the decline in optimal target accuracies in Fig. 4c. Additionally, it can be observed that the optimal learning rate values – shown in cyan markers in Fig. 4b – inversely decrease with the increase in device cycle-to-cycle variability. This suggests that in order to optimally train a ReRAM-based hardware network, smaller gradient steps proportional to the cycle-to-cycle variability of the underlying ReRAM devices may be required. Fig. 4a also highlights the importance of performing a learning rate optimization. This is because modeling bias results at unoptimized learning rates can be potentially misleading.

Finally, network simulations of binning models show that the network modeling bias tends to increase with the increase in noise, as captured in Fig. 4d. It can also be observed that the network results are largely unreliable in terms of the sign of modeling bias – with some cases over-estimating target accuracy, and others under-estimating target accuracy. Overall, we conclude that the binning algorithm fails to effectively capture target mean and standard deviation profiles when the standard deviation in the input dataset is high given a fixed number of data points, i.e., beyond 2 nS given 4,000 $(G, \Delta G)$ data points.

*C. Impact of Non-Linearity*

The impact of the device non-linearity on modeling bias is investigated using variants of the analytical model at different device non-linearity values. The chosen parametrization for non-linearity is formulated as follows: given a base jump table model with mean profile $\mu(G)$, the corresponding model at non-linearity k – where k is a positive integer – is the model with mean profile $k\mu(G) = \{k\mu_1, k\mu_2, ... k\mu_n\}$. Note that the standard deviation profile is unaltered, though the multiplication of the mean profile by a constant can lead to a change in the net coefficient of variation (defined as dispersion around the mean). Fig. 5 summarizes our analysis on device non-linearity.

Fig. 5a shows the conductance trajectories of target and binning models at different non-linearities. Contrary to Fig. 4a where target devices get noisier, here we see that target devices show less noise under our chosen scheme of increasing non-linearity. This is due to the net coefficient of variation decreasing with the increase in device non-linearity. Additionally, it can be observed that binning modelling is able to better model target devices with high non-linearity and low cycle-to-cycle variability, compared to target devices that have low non-linearity and high cycle-to-cycle variability. This improvement is manifested in the OVLE and SSD device metrics shown in Fig. 5b, which are consistently significantly lower compared to devices from Fig. 3c. As the non-linearity increases, we make the following two observations at the network level. Firstly, the optimal accuracy of target models increases. Secondly, the modeling bias performance of corresponding binning models improves, as captured in the thinning gray region in Fig. 5c. The convergence curves in Fig.

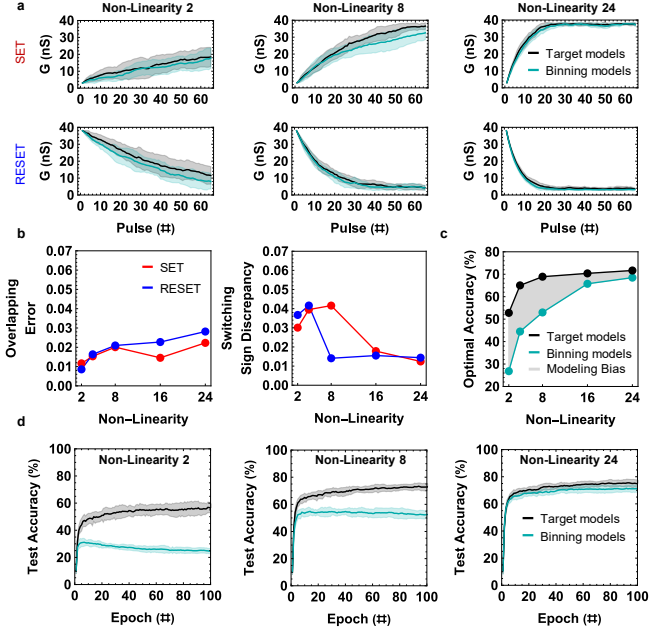

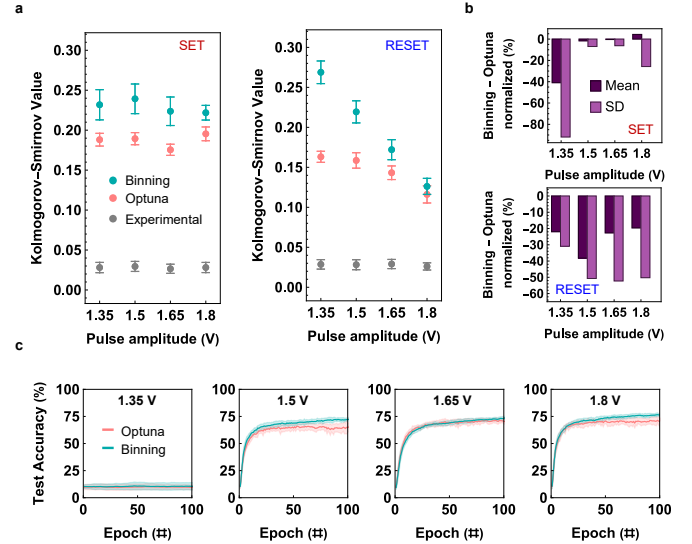

**Fig. 5.** Impact of device non-linearity on device modelling and respective network simulations. (a) Conductance trajectories of target and binning jump table models with increasing non-linearity for up to 64 applied pulses. At higher non-linearity, devices saturate faster. (b) Overlapping error and switching sign discrepancy of the binning models. (c) Comparative optimal accuracies at different levels of device non-linearity. (d) Corresponding optimized network training curves at different levels of non-linearity.

5d reinforce these observations. Our device metrics do not seem to fully capture modeling bias characteristics, contrary to Fig. 3 where target profiles were unaltered. We hypothesize that this happens because of significant changes to switching behavior from one non-linearity model to another, and acknowledge that our device metrics need to be further refined to better capture device switching intricacies.

*D. Experimental Model Verification*

While prior sections investigated synthetic datasets, this section summarizes the modeling results on experimental data. We use our experimental datasets (Fig. 1c, d) to derive jump table models using the approach described in Section II D.3.

Fig. 6a shows Kolmogorov-Smirnov test error values for the different SET and RESET datasets for the binning and Optuna models. To illustrate the room for optimization in these models, we also plot the test error between the experimental model and experimental test sub-sets. These values are the lowest as expected, as the two are indeed drawn from exactly the same distribution. The subsampling of the experimental datasets, the binning modelling, and the error calculations were repeated for a total of 20 iterations for maintaining statistical significance. An error bar indicates one standard deviation.

There is a performance gap between what the binning models can do to approximate the experimental data, as observed by the gap between the cyan and gray points in Fig. 6a. This is line with the results from Fig. 4. Our proposed Optuna-based algorithm for optimization produces (G, ΔG) data is significantly closer to the experimental reality compared to the binning models in all investigated cases, with lower K-S values. Since Optuna models still exhibit a performance gap from the experimental data, better interpolation methods still need to be explored. Nevertheless, network simulations of the Optuna models seem to represent experimental reality closer than corresponding binning models. End-to-end experimental verification is left for future work.

**Fig. 6.** Experimental Model Verification. (a) Kolmogorov-Smirnov test errors of the experimental test dataset with the experimental model dataset and synthesized datasets from the binning and the Optuna-optimized binning models. (b) Normalized difference (%) between mean and standard deviation profiles of the binning models with the Optuna models. (c) Optimized network training curves investigated pulse amplitudes.

Fig. 6b shows the normalized difference of the mean and standard deviation profiles between binning and Optuna models. In all cases, it can be observed that differences in the standard deviation profile are higher than the differences in the mean profile. A negative bar links to a binning profile underestimating the corresponding Optuna profile, and it can be seen that binning tends to generally underestimate the standard deviation. This underestimation of the device noise directly manifests in the network convergence curves shown in Fig. 6c, where binning models over-promise the test accuracy that would be obtained by a ReRAM network implemented using our devices compared to Optuna models.

IV. DISCUSSION

These results point to the need to understand the goodness of a device model in connection with the experimental reality of ReRAM hardware for in-memory compute and neural network accelerators. In hardware implementations, the obtained accuracy depends on the underlying distribution of the device population used for implementation. While this underlying distribution is unknown and can only be approximated indirectly via electrical measurements, a good statistical model of the device based on these measurements should be able to predict the target accuracy that a hardware prototype would

achieve – if available. The value of the accuracy is not of critical consideration here per se; the modeling bias, defined as the difference between simulated accuracy and target accuracy, is more relevant. For example, a software simulation based on a given device model may achieve an accuracy higher than the target model, supposing such a hardware model exists for comparison. This outcome would be undesirable since the simulation results overpromise on the particular device technology proposed. For our investigation at the network level, we are only interested in how well a statistical model can approximate the underlying target distribution of the device population, which does not have to be the model that achieves the highest network training accuracy.

To support these co-design efforts in the absence of network experimental data, this work has proposed the use of automatic optimizers such as Optuna to estimate network convergence as the benchmark for estimating the modeling bias performance of corresponding experimentally derived device models. While this approach does not replace the need of experimentally obtaining network performance estimates and studying exact device performance, it could be used as a short-term solution to support iterative device and network co-optimization, particularly when experimental results of network training might be difficult to obtain.

We have shown that there are limitations to binning models in approximating synthetic target model convergence. Specifically, binning models exhibit poor device modeling and a high modeling bias when a) the input dataset is scarce – having fewer than 4,000 data points, and b) when the device cycle-to-cycle variability is high – beyond 2 nS in our chosen network training regime. These limitations mean that exploring other interpolation methods in the context of jump table-based device modelling is promising. Our prior work in device modeling with Gaussian Process Regression methods, such as ordinary Kriging [26] hints at a potential method that will be investigated in the future for network training.

It is important to point out that this initial investigation is restricted to Gaussian interpolation methods applied mostly to linear mean and constant standard deviation profiles. In reality, emerging devices can follow various profiles, e.g. piecewise linear profile [20], parabolic mean and constant standard deviation [19], etc. Moreover, these Gaussian interpolation methods can be insufficient if the underlying data is skewed due to physical bounds and limitations based on the parasitic resistance of the device. More general interpolation methods, e.g., based on skew normal distribution, could better predict complex device behavior and possibly neural network behavior more accurately and consistently. We plan to explore in future work how various interpolation modeling methods compare to one another in terms of corresponding network modeling bias where the underlying distributions are complex and non-trivial.

Since our device metrics and modeling bias results both monotonically decrease (e.g., Fig. 3), we hypothesize that these metrics could be the first step towards having a mechanism to predict modeling bias performance of different device models relative to one another. However, before such a link could be established, these metrics need to be further investigated and refined to fully understand their connection with modeling bias. Network parameters such as the learning rate, loss function, and network dimensionality would have to be studied in conjunction with device training statistics such as the number of times SET and RESET tables are used over training, number of SET or RESET pulses applied, etc. This is left for future work.

## V. Conclusions

This paper proposes the concept of "modeling bias" as a useful metric to quantify the goodness of a device model at the neural network level. To exemplify this concept, the binning interpolation method was used for modeling ReRAM device jump tables and its applicability in predicting neural network convergence behavior was investigated. For testing, a wide range of synthetic Gaussian datasets with linear mean and standard deviation profiles as well as experimentally obtained datasets was used, in conjunction with metrics at the device and network level were explored. The results show that device models based on binning can lead to unreliable modeling bias behavior, sometimes over-promising and sometimes under-promising the network accuracy. Better interpolation methods that have lower bias than binning, particularly at high device switching noise and low number of data points will be investigated in the future. Promising device metrics seem to have a similar trend with modeling bias at the network level, but additional investigations are needed for more complex device switching profiles and more difficult image classification datasets. Additionally, simulated accuracy results will be experimentally verified. Finally, the proposed Optuna algorithm for device modelling will be further refined by exploring other, more application-friendly error metrics beyond the Kolmogorov-Smirnov test. This work highlights the need for additional correlated efforts in device / network modeling and prototyping for in-memory compute and neural networks.

## Acknowledgment

We thank Mark Anders and Lin Wang for prior useful discussions. The authors acknowledge the use of high-performance computing clusters, advanced support from the research technology services, and IT support at The George Washington University and NIST.

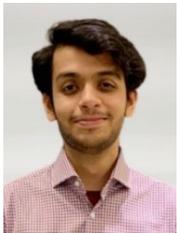
**Osama Yousuf** (Student Member, IEEE) received a B.S. degree in computer science with a minor in mathematics from Habib University, Karachi, Pakistan in 2020. He is currently pursuing a Ph.D. degree in computer engineering at the George Washington University, Washington, DC, USA in the Adaptive Devices And Microsystems (ADAM) group as a research assistant. He also works as a research associate at the National Institute of Standards and Technology (NIST), Gaithersburg, MD. His research interests include robust neural network training algorithms, modelling emerging non-volatile memory devices, and prototyping ReRAM-based accelerators.

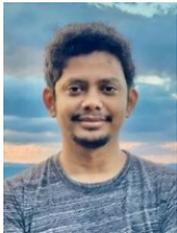
**Imtiaz Hossen** (Student Member, IEEE) received the B.Sc. degree in electrical and electronic engineering from the University of Dhaka, Dhaka, Bangladesh, in 2016, and the M.Sc. degree in electrical and computer engineering from Marquette University, Milwaukee, WI, USA, in 2020. He is currently pursuing a Ph.D. degree in electrical engineering at the George Washington University, Washington, DC, USA in the Adaptive Devices And Microsystems (ADAM) group as a research assistant. His research interests are nano-scale fabrication and testing of novel memory devices, device modeling, heterogeneous integration of ReRAM in CMOS foundry chips, ultra-high resolution temperature sensors and RF/microwave sensors.

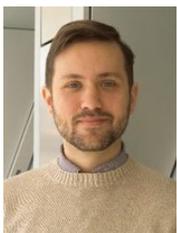
**Matthew W. Daniels** is a research physicist in the Alternative Computing Group at the National Institute of Standards and Technology (NIST), Gaithersburg, MD. He received a B.S. in physics with a minor in mathematics from Clemson University in 2012 and a Ph.D. in physics from Carnegie Mellon University in 2017. His dissertation work was on topological physics and theoretical antiferromagnetic spintronics. His current research program seeks to build a theory of hardware neural networks and neuromorphic systems. He currently works on understanding how time and stochasticity can be effectively utilized as primitive encodings for computing systems and on designing digital architectures that use the spectrally-biased, low-rank properties of neural network gradients to accelerate their training in hardware.

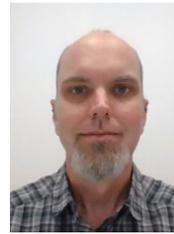
**Martin Lueker-Boden** is an engineering research director for Western Digital Corporation (WDC). He earned a Ph.D. in physics from the University of California at Berkeley in 2010. After earning his doctorate, he worked as a post-doctoral scholar at the California Institute of Technology. He started working at WDC in 2014 with a research focus on developing novel applications for non-volatile memory devices, including development of new cell technologies, methods for media management techniques, security considerations, I/O frameworks, software optimizations and in-memory/neuromorphic computing. His work prior to joining WDC focused on the development of superconducting low-noise microwave radiation sensors, and the statistical analysis of extra-galactic radiometry data.

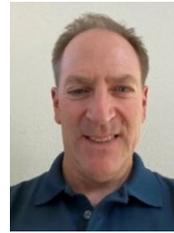
**Andrew Dienstfrey** received the Ph.D. in Applied Mathematics from the Courant Institute of Mathematical Sciences at New York University. Dienstfrey joined the Mathematical and Computational Sciences Division at the National Institute of Standards and Technology (NIST), Boulder, CO in 2000. His research interests include applications of mathematics to computational physics and numerical analysis.

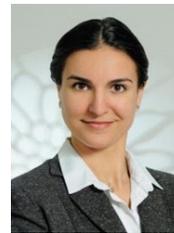
**Gina C. Adam** (Senior Member, IEEE) received the B.Sc. degree in applied electronics from the University Politehnica of Bucharest, Bucharest, Romania, in 2010, and the Ph.D. degree in electrical and computer engineering from the University of California at Santa Barbara, CA, USA, in 2015. From 2016 to 2018, she was a Research Scientist with the National Institute for Research and Development in Microtechnologies, Voluntari, Romania, and a Visiting Scholar with the École Polytechnique Fédérale de Lausanne, Lausanne, Switzerland. She is currently an Assistant Professor of electrical and computer engineering with the School of Engineering and Applied Science, George Washington University, Washington, DC, USA. Her current research interests include resistive switching devices and their use in memory storage, computing, and communications applications.